\documentclass[aip,reprint]{revtex4-1}

\draft 

\usepackage{graphicx}
\usepackage{dcolumn}
\usepackage{bm}

\usepackage[utf8]{inputenc}
\usepackage[T1]{fontenc}
\usepackage{mathptmx}
\usepackage{etoolbox}
\usepackage{upgreek}

\makeatletter
\def\@email#1#2{%
 \endgroup
 \patchcmd{\titleblock@produce}
  {\frontmatter@RRAPformat}
  {\frontmatter@RRAPformat{\produce@RRAP{*#1\href{mailto:#2}{#2}}}\frontmatter@RRAPformat}
  {}{}
}%
\makeatother

\begin{document}


\title{Coherent control of terahertz-scale spin resonances using optical~spin--orbit~torques}

\author{Julian Hintermayr}
\email[Electronic mail: ]{j.hintermayr@tue.nl}
\affiliation{%
 Department of Applied Physics, Eindhoven University of Technology,\\
 P.O. Box 13, 5600 MB Eindhoven, the Netherlands
}%
\author{Paul M. P. van Kuppevelt}
\affiliation{%
 Department of Applied Physics, Eindhoven University of Technology,\\
 P.O. Box 13, 5600 MB Eindhoven, the Netherlands
}%
\author{Bert Koopmans}
\affiliation{%
 Department of Applied Physics, Eindhoven University of Technology,\\
 P.O. Box 13, 5600 MB Eindhoven, the Netherlands
}%

\date{\today}

\begin{abstract}
Using optically generated spin--orbit torques induced by the heavy metal Pt, we demonstrate coherent control of GHz ferromagnetic resonances in Pt/Co/Pt multilayers as well as sub-THz exchange resonances in [Gd/Co]$_2$ multilayers. Employing a double-pump setup, we show that depending on the helicities of the pump pulses, spin resonances can either be coherently amplified or suppressed if the time delay between the arrival of the pump beams is chosen appropriately. Furthermore, investigating phase and amplitude of the exchange-driven modes, we identify features that challenge the current understanding of optically generated spin--orbit torques, and we discuss possible explanations.
\end{abstract}

\maketitle

\section{Introduction}
The ultrafast manipulation of spins on the nanoscale poses one of the core challenges in spintronics~\cite{Kirilyuk:2010, Walowski:2016}. Recent breakthrough experiments have revealed the potential of optically generating spin currents in non-magnetic heavy metals (HMs) using circularly polarized light, which can be injected into neighbouring ferromagnetic (FM) layers, exerting an ultrashort spin-transfer torque~\cite{Choi:2017, Choi:2020}. This novel phenomenon was termed optical spin--orbit torque (OSOT), as its fundamental prerequisite is spin--orbit coupling in the layer generating the spin photocurrent. Under laser excitation with normal incidence, the spin polarization points out-of-plane (OOP), with its direction reversible by altering the circular polarization of the trigger~\cite{Choi:2017, Choi:2020, Iihama:2022}. This clearly distinguishes the OSOT from its electrical counterpart, known as spin--orbit torque (SOT), in HM/FM bilayers, where the symmetry of the spin--Hall effect typically predicts an in-plane polarization of the injected spin current. Additionally, while a breaking of inversion symmetry is necessary for SOTs, it is not a requirement for OSOTs. Finally, while time integrated effects on the magnetization orientation using optical excitation are typically quite modest, the torque applied during the optical pulse can be very large.

In comparison to other \textit{optical} spin excitation methods, such as ultrafast demagnetization of neighboring layers, OSOTs offer additional advantages. In demagnetization-induced spin currents, the polarization is predetermined by the magnetization direction of said layer and its sign is not easily reversible~\cite{Malinowski:2008, Schellekens:2014b, vanHees:2020, Beens:2020}. Further studies have shown that this excitation mechanism can not only be used to trigger ferromagnetic resonance (FMR) modes in the neighboring FM layer, but also quantum-confined standing spin waves with frequencies of up to 1~THz~\cite{Choi:2020b}, making them a useful tool for ultrafast spintronics and magnonics.

This study explores the possibility of coherently controlling spin resonance modes using multiple pump beams with adjustable delays and polarization states. Employing time-resolved magneto-optical techniques, we demonstrate that FMR modes in Pt/Co/Pt can be triggered by the first pump pulse, with the second pulse amplifying or suppressing the mode based on its delay and polarization. Extending this concept, we find that \textit{ferrimagnetic} (FiM) exchange resonance modes (EXMs) in Co/Gd-based systems—offering much higher frequencies (sub-THz) and obviating the need for external fields—can likewise be manipulated through this method. Furthermore, carefully analyzing phase and amplitude of the oscillatory EXM signal, we identify characteristics in the dynamics of the Co sublattice that contradict the current understanding of OSOTs and we propose an excitation scheme based on efficient absorption of angular momentum by Gd to explain our findings. The above insights on the coherent control of terahertz-scale EXMs hold great promise for the advancement of ultrafast spintronic computation devices.

\section{Magnetic resonance modes}
We start by briefly discussing the types of spin resonances that are allowed in the FM and FiM samples used in this study. In FM thin films exposed to a magnetic field of magnitude $H$ pointing along an in-plane (IP) direction, an impulsive excitation from the equilibrium state results in a precession around said field, where the frequency of the oscillation is given by the Kittel equation $f=\gamma\mu_0 \sqrt{H(H+M_\mathrm{eff})}/(2\pi)$. $M_\mathrm{eff}$ denotes the effective magnetization, $\gamma$ the gyromagnetic ratio and $\mu_0$ the vacuum permeability. In ferrimagnetic materials with multiple antiferromagnetically coupled sublattices on the other hand, built-in exchange fields enable field-free high-frequency EXMs~\cite{Keffer:1952}. For simplicity's sake, we consider the case of a Co-Gd alloy where each sublattice can be treated as a single macrospin. While the multilayered structure of our [Co/Gd]$_2$ synthetic FiM in principle permits higher order modes, we find that the description in terms of an alloy adequately accounts for all experimental findings in this study. The prerequisite for EXMs to exist is a canted state between the two sublattices, which can for instance be triggered by ultrafast spin injection~\cite{Hintermayr:2023} or intense THz pulses~\cite{Blank:2021}. A schematic of the local arrangements of Co and Gd magnetizations $\mathbf{M}_\mathrm{Co,Gd}$ during the EXM is given in Fig.~\ref{fig:schematic}~\textbf{c}. Furthermore, canting angles $\theta_\mathrm{Co}\neq \theta_\mathrm{Gd}$ and directions of exchange fields $\mathbf{H}_\mathrm{ex}^\mathrm{Gd,Co}$ are indicated. The frequency of the exchange mode is given by $f=\mu_0\lambda\gamma_\mathrm{Co}\gamma_\mathrm{Gd}\left(\frac{M_\mathrm{Co}}{\gamma_\mathrm{Co}} - \frac{M_\mathrm{Gd}}{\gamma_\mathrm{Gd}} \right)/(2\pi)$ with $\lambda$ the Weiss constant. We note that $\lambda$ is reduced in the multilayer case compared to a true alloy, due to reduced exchange coupling of Co and Gd which only takes place across interfaces. The dispersion relation implies that the resonance frequency depends on the degree of angular momentum compensation and can be tuned by varying the ratio of Co and Gd moments in the sample, which has been demonstrated in Co-Gd based alloy~\cite{Stanciu:2006, Mekonnen:2011, Haltz:2022} and multilayer samples~\cite{Hintermayr:2023}.

\section{Results and discussion}
\begin{figure}[htbp]
    \centering
    \includegraphics[width=8.6cm]{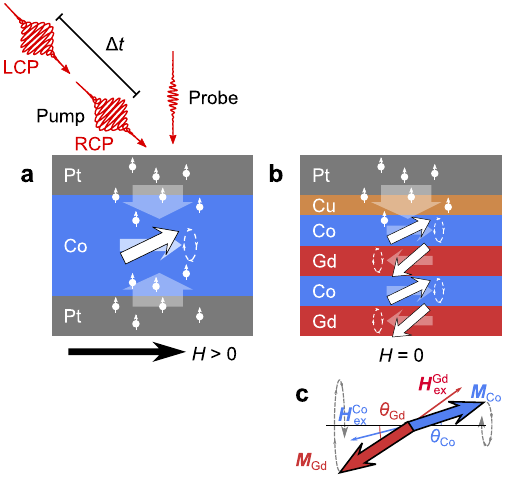}
    \caption{\textbf{a}~Schematic representation of the experimental setup, displaying the double pump-probe measurement technique with variable helicities of the two pump beams exciting ferromagnetic resonance spin dynamics in a Pt/Co/Pt multilayer. \textbf{b}~Sample geometry of the Co/Gd-based multilayer, displaying field-free exchange-driven spin resonances. \textbf{c}~Local orientations of canted Co and Gd magnetic moments and exchange fields, facilitating the exchange resonance mode.}
    \label{fig:schematic}
\end{figure}

We prepare magnetic thin film samples by dc magnetron deposition at room temperature with a base pressure of the system below $1\cdot 10^{-8}$~mbar. The two films investigated in this study include the FM/HM heterostructure sub/Ta(4~nm)/Pt(4~nm)/Co(5~nm)/Pt(4~nm) and the synthetic ferrimagnetic FiM/HM heterostructure sub/Ta(4~nm)/[Gd($t_\mathrm{Gd}$)/Co(1.2~nm)]$_2$/Cu(1~nm)/ Pt(4~nm), in which the Gd thickness $t_\mathrm{Gd}$ is wedged ranging from 0--3~nm and sub refers to Si/SiO$_2$(100~nm). Schematics of the two sample stacks are depicted in Fig.~\ref{fig:schematic}~\textbf{a} and \textbf{b}. The samples are used to investigate FMR and EXM resonances, respectively.

We measure the static magnetic properties using the longitudinal magneto-optic Kerr effect (L-MOKE) to investigate the IP magnetization reversal behaviour of the films. The magnetization dynamics are captured using the time-resolved magneto-optic Kerr effect (TR-MOKE) in polar geometry using  $\sim 100$~fs laser pulses with a central wavelength of 780~nm and a repetition rate of 80~MHz. At this wavelength, the magneto-optic signal of Gd is very weak~\cite{Erskine:1973}, the measured dynamics therefore describe the precessions of only the Co sublattice. Typical pump fluences are in the range of 1.0~mJ~cm$^{-2}$ with pump and probe sizes of 16~$\upmu\mathrm{m}$ and 8~$\upmu\mathrm{m}$ respectively. A standard double modulation technique is applied to improve the signal recovery. Spin dynamics are excited by one or two pump pulses with variable delay and helicity, schematically shown in Fig.~\ref{fig:schematic}~\textbf{a}.

\begin{figure}[htbp]
    \centering
    \includegraphics[width=8.6cm]{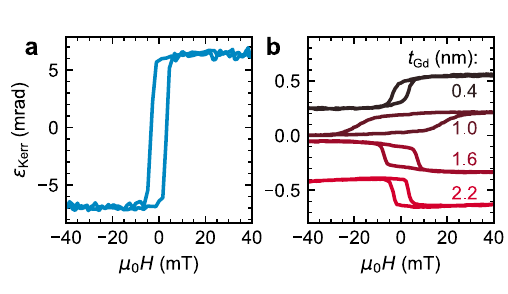}
    \caption{MOKE hysteresis loops in longitudinal geometry of \textbf{a},~the Pt(4~nm)/Co(5~nm)/Pt(4~nm) sample and \textbf{b},~the [Co(1.2~nm)/Gd($t_\mathrm{Gd}$)]$_2$ multilayer sample for selected values of $t_\mathrm{Gd}$, as indicated in the Figure. The change in sign of the hysteresis loop indicates the magnetic compensation point.}
    \label{fig:LMOKE}
\end{figure}

\begin{figure}[htbp]
    \centering
    \includegraphics[width=7cm]{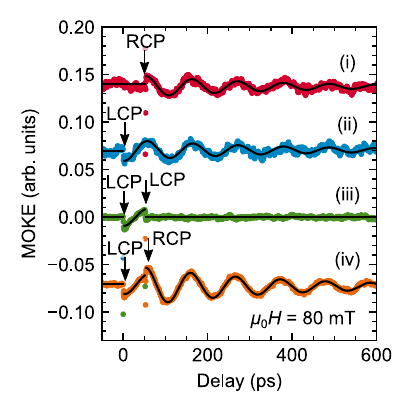}
    \caption{Oscillation measurements of 9.4~GHz FMR dynamics in a Pt(4~nm)/Co(5~nm)/Pt(4~nm) multilayer triggered by one (trace (i), (ii)) or two (trace (iii), (iv)) circularly polarized pump pulses with a fixed delay of half an oscillation period. Traces (i) and (ii) display the behaviour of single pump beams with opposite helicities. Trace (iii) shows the scenario where the resonance mode triggered by the first pump beam is annihilated by a second pump beam with the same helicity. The case of opposite helicities, resulting in an amplification of the signal, is shown in trace (iv). The applied field is 80~mT. Vertical offsets are for clarity.}
    \label{fig:FMR}
\end{figure}

We first turn our attention towards the ferromagnetic Pt/Co/Pt film. Due to the high Co thickness, the magnetic shape anisotropy outweighs the interfacial anisotropy arising from Pt/Co interfaces, leading to a film whose magnetization lies fully IP. This is confirmed by the L-MOKE hysteresis loop shown in Fig.~\ref{fig:LMOKE}~\textbf{a}, clearly exhibiting easy-axis behaviour along the IP direction. In Fig.~\ref{fig:FMR}, time traces of magnetization dynamics induced by one or two circularly polarized pump pulses are depicted. Solid lines in the Figure represent piecewise fits of damped cosine functions. Outlying data points are observed when the probe beam overlaps with either one of the pump beams, resulting from coherence effects and representing a non-magnetic artifact~\cite{Eichler:1984, Luo:2009}. Trace (i) shows an FMR oscillation with a frequency of 9.4~GHz that is triggered by a RCP laser pulse with a time delay of half an oscillation period with respect to zero time delay. The effective damping parameter of the mode calculates as $\alpha_\mathrm{eff}=1/(2\pi f\tau)\approx 0.05$. Comparing this result to the blue trace, which is excited by a LCP laser pulse showing a negative amplitude, we confirm that reversing the helicity indeed inverts the sign of the spin current pulse that triggers the dynamics. The phases of the two oscillations of different helicity show a shift of $-6\pm 3^\circ$ and $165\pm 3^\circ$ for RCP and LCP excitations with respect to the arrival of the pump beam, respectively. It can therefore be concluded that the dynamics is driven predominantly by a damping-like torque, suggesting that the injection of spin photocurrents generated by Pt is a likely mechanism, in agreement with existing literature. Whereas the Ta seed layer is also known to give rise to weak spin photocurrents, the efficiency of Pt is around five times stronger~\cite{Choi:2020}. Since, in addition, the Ta layer is not directly interfaced with Co and the thickness of Pt is roughly equal to its spin diffusion length, we conclude that the dynamics are indeed driven by spin currents generated in Pt. We note that other mechanisms like the inverse Faraday effect~\cite{Freimuth:2016} and the optical Rashba-Edelstein effect~\cite{Iihama:2022} can give a field-like contribution to the torque, possibly explaining the deviations from $0^\circ$ and $180^\circ$ for RCP and LCP excitations.

Pumping the sample with two laser pulses of same helicity and a time delay of half an oscillation period, we find that the oscillation can be completely halted again, shown as trace (iii) in Fig.~\ref{fig:FMR}. Inverting the circular polarization of the second pump, the amplitude of the mode is doubled, shown as trace (iv) in Fig.~\ref{fig:FMR}. Furthermore, no significant change in the phase of the oscillation is observed. Thus, the OSOT acts on the magnetization vector independently of whether the vector lies fully in the plane or describes a small-angle precession, at least in the limit of oscillation periods that are long compared to the timescale of the torque.

Having shown that OSOTs allow for coherent control of GHz FMR modes in FMs, we turn our attention to the Co/Gd-based FiM multilayers hosting higher frequency EXM dynamics. A schematic representation of the sample architecture is shown in Fig.~\ref{fig:schematic}~\textbf{b}. As in the experiment above, the Pt layer functions as the spin generation layer. A thin Cu layer that is virtually transparent to spin currents separates the HM from the FiM to avoid interactions between Co and Pt that could lead to perpendicular magnetic anisotropy close to the magnetic compensation point, where demagnetizing effects are weakest. To prove that the sample indeed shows an IP easy axis and that achieving compensation using the given sample design is possible, we measure L-MOKE hysteresis loops on the sample at points with different Gd thicknesses. Loops for four exemplary compositions are shown in Fig.~\ref{fig:LMOKE}~\textbf{b}. All curves exhibit easy axis characteristics, while a reversal of the sign of the hysteresis loop is observed between 1.0 and 1.6~nm Gd. Furthermore, the coercive field is largest around those values, which is a characteristic of the magnetization compensation point. Note that the \textit{angular momentum} compensation point at which the frequency of the EXM softens lies at slightly lower values of $t_\mathrm{Gd}$ than the magnetization compensation point, since $\gamma_\mathrm{Co}> \gamma_\mathrm{Gd}$.

Before discussing experimental results, it is worth briefly revising spin injection effects in FiMs that are distinctly different from effects in FMs. Whereas FMs in an excited state will oscillate around the effective magnetic field, spin-injected FiMs will experience torques from both the effective field as well as exchange torques, the latter of which will cause precessions about the total angular momentum vector. Precession amplitudes then depend not only on (element-specific) canting angles induced by the OSOT but also on the degree of angular momentum compensation. Further details can be found in Ref.~\cite{Hintermayr:2023}.

\begin{figure}[htbp]
    \centering
    \includegraphics[width=7cm]{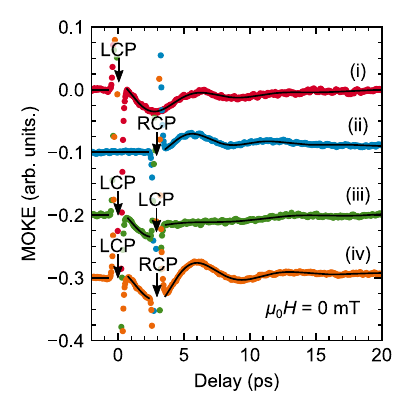}
    \caption{The same experiment as described in Fig.~\ref{fig:FMR} performed on a [Gd(1.8~nm)/Co(1.2~nm)]$_2$/Cu(1~nm)/Pt(4~nm) multilayer, displaying a field-free EXM frequency of $\sim 0.15$~THz in the absence of a magnetic field.}
    \label{fig:EXM}
\end{figure}

We choose a [Co/Gd]$_2$ multilayer with a Gd thickness of 1.8~nm for the second study, providing EXM frequencies in the sub-THz range while maintaining a sizable oscillation amplitude. Further details on the relation between EXM amplitude and frequency in the context of ultrafast spin injection experiments are provided in Ref.~\cite{Hintermayr:2023}. Experimental results on single and double-pump excitation in Co/Gd-based FiM multilayers are presented in Fig.~\ref{fig:EXM}. The magnetic responses to pumping with left and right circularly polarized laser pulses are plotted as traces (i) and (ii), respectively. Note that in the RCP case, the pump is again delayed by half an oscillation period. As in the FM case, reversing the pump circularity inverts the oscillation amplitude. To obtain satisfactory fits to the EXM traces, an exponentially decaying background is added to the fitting functions. A discussion on why this modification is necessary is provided further below. The frequency of the precession mode extracted from the fit is 0.15~THz and exhibits a strong effective damping of $\alpha_\mathrm{eff}\approx 0.25$. Such high damping values are expected for EXMs close to the angular momentum compensation point, where a characteristic maximum is theoretically predicted~\cite{Schlickeiser:2012}.

Double pump excitation with two LCP pulses leads to an almost complete suppression of the EXM, as shown as trace~(iii) in Fig.~\ref{fig:EXM}. The small but finite remanent amplitude could be explained by the high effective damping. By the time the second pump beam, exhibiting the same power as the first one, impinges on the sample, the oscillation has already undergone substantial damping. The angular momentum injected by the second pump therefore overcompensates the amplitude of the first oscillation, leading to a finite leftover amplitude. When pumping with opposite helicities, plotted as trace (iv), the resulting precession amplitude is again about twice the amplitude of the one pumped by a single pulse. We therefore conclude that OSOTs also offer coherent control over sub-THz EXMs in FiMs.

\begin{figure}[htbp]
    \centering
    \includegraphics[width=8.6cm]{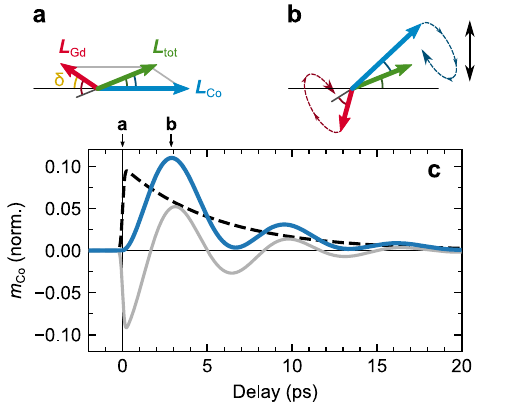}
    \caption{Scheme to explain the observed EXM dynamics in the [Gd/Co]$_2$ sample. \textbf{a},~Cartoon of the spin current excitation in the scenario where only Gd absorbs angular momentum. \textbf{b},~Precessional state after half an oscillation period when the deflection of the Co moment is largest. \textbf{c},~Time dependency of the z-component of the Co magnetization $m_\mathrm{Co}$ (blue) following the excitation mechanism in \textbf{a}. The offset due to the canting of the total angular momentum vector is shown as  the dashed line, whereas the oscillatory behavior of the Co moment around $\mathbf{L}_\mathrm{tot}$ is shown as the grey line. The points in time at which scenarios \textbf{a} and \textbf{b} occur are indicated on the top x-axis.}
    \label{fig:Gd_abs}
\end{figure}

A peculiarity of the EXM traces is that after the laser trigger, the MOKE signal starts oscillating from its initial IP orientation, which is in contrast to the FMR measurements, where the spin photocurrent excitation resulted in a quasi-instantaneous canting, at least compared to typical timescales of the precession period. Note that nevertheless, the phase of the oscillation of the Co moment is again close to 0$^\circ$, meaning the oscillation starts at its maximum amplitude. However, a significant background on top of the oscillatory signal is now present. While a helicity-dependent background signal that is not related to the canting of the Co magnetization could explain the measured traces, this signal would need to be of exactly the same magnitude as the canting induced in the Co, which seems unlikely.

In the following, we present an alternative attempt at explaining the observed behavior, following an excitation scheme that proposes significant absorption of angular momentum in the Gd sublattice. This assumption is required, since absorption of angular momentum by Co on ultrafast timescales would lead to a deflection of the Co moment, which in turn would manifest itself in an offset in the magneto-optic signal shortly after zero time delay. This notion, however, contradicts our measurements. The proposed behavior is described in Fig.~\ref{fig:Gd_abs}. Assuming absorption of angular momentum only by Gd, its total angular momentum vector $\mathbf{L}_\mathrm{Gd}$ is tilted by an angle $\delta$ and acquires an OOP component, schematically shown in Fig.~\ref{fig:Gd_abs}~\textbf{a}. As a result, the total angular momentum vector $\mathbf{L}_\mathrm{tot}$, which is the sum of the sublattice angular momenta $\mathbf{L}_\mathrm{Co,Gd}$ is also partially tilted OOP. Since an angle between $\mathbf{L}_\mathrm{Co}$ and $\mathbf{L}_\mathrm{Gd}$ arises, $\mathbf{L}_\mathrm{Co}$ experiences an exchange torque leading to an oscillation of both sublattices around $\mathbf{L}_\mathrm{tot}$ with the same rotational sense. Due to the canting of $\mathbf{L}_\mathrm{tot}$, the precession of the z-component of $\mathbf{L}_\mathrm{Co}$ is offset from zero by half an oscillation amplitude in z-direction. The precessional state after half a period is shown in Fig.~\ref{fig:Gd_abs}~\textbf{b}. At this point, the Co moment exhibits the strongest deflection with respect to the sample plane. The strong effective damping of the EXM leads to a rapid attenuation of both the precession amplitude, as well as the OOP canting of $\mathbf{L}_\mathrm{tot}$. A possible trajectory of the z-component of Co following the excitation mechanism described above is shown in Fig.~\ref{fig:Gd_abs}~\textbf{c} as a blue line. The dashed black line indicates the decaying background due to the canting of $\mathbf{L}_\mathrm{tot}$ out of the film plane. The grey line shows the isolated Co precession without the background, indicating that the phase of the oscillation is indeed equal to 0$^\circ$.


Whereas the mechanism discussed in the paragraph above suffices in explaining the measured dynamics, we note that it is in contrast to existing literature, which reports that transverse spin currents are efficiently absorbed in Co with a characteristic decay length of $\sim0.9~$nm~\cite{Lalieu:2017b}, implying that most of the injected spin current should be absorbed by Co. Yet, we want to stress two differences between experiments in Ref.~\cite{Lalieu:2017b} and our work. First, these experiments were performed on elemental Co layers and second, the ultrafast spin current is generated by laser-exciting a neighboring ferromagnetic layer instead of a nonmagnetic HM. While the exact mechanism behind this phenomenon is still under debate~\cite{Battiato:2010, Battiato:2012, Beens:2020, Lichtenberg:2022, Gupta:2023}, the energies of the mobile spin-polarized electrons involved in the process are likely at different levels above the Fermi energy compared to the mechanism used in this work. This could result in differences in spin diffusion length and spin absorption. Furthermore, the optical generation of \textit{orbital} angular momentum currents by circularly polarized laser pulses in Pt~\cite{Hamamera:2024} and across Co/Cu interfaces~\cite{Busch:2023}, constituting part of our sample stack, have been theorized very recently. Yet, experimental evidence of this effect is still lacking to the best of our knowledge, and the details of the interactions between such currents and our specific sample are not clear. On a similar note, the role of orbital-to-spin conversion in $4f$ elements has also been addressed~\cite{Sala:2022}. If such currents were more efficiently absorbed by Gd than by Co, they would serve as a sufficient explanation for the observed behavior. Since the main goal of this study is the investigation of coherent control by double pumping, further investigations on the details of the generation, transport, and absorption of angular momentum currents in the non-trivial multilayers used in our experiments are beyond the scope of this work. Nonetheless, these aspects present intriguing avenues for future research. 


\section{Conclusion}
In conclusion, we have shown that GHz FMR modes and sub-THz EXM precessions can be excited by optical spin--orbit torques originating from Pt layers in Co and Co/Gd-based multilayers, respectively. Further, consecutive spin injection may amplify the amplitude of the oscillation or bring the system back to the ground state, depending on the circular polarization of the second pump pulse. Finally, we have observed an intriguing feature in the magnetization traces of EXMs excited in Co/Gd-based multilayers, which could be explained by a strong absorption of angular momentum by Gd instead of Co. Our findings hold great promise for the future of high-speed spintronic computation devices.

\section*{Data availability}
The data that support the findings of this study are available from the corresponding author upon reasonable request.

\section*{Conflict of Interest}
The authors have no conflicts to disclose.

\section*{Acknowledgments}
This project has received funding from the European Union’s Horizon 2020 research and innovation programme under the Marie Skłodowska-Curie grant agreement No 861300.

\bibliography{Ref}

\begin{thebibliography}{29}%
\makeatletter
\providecommand \@ifxundefined [1]{%
 \@ifx{#1\undefined}
}%
\providecommand \@ifnum [1]{%
 \ifnum #1\expandafter \@firstoftwo
 \else \expandafter \@secondoftwo
 \fi
}%
\providecommand \@ifx [1]{%
 \ifx #1\expandafter \@firstoftwo
 \else \expandafter \@secondoftwo
 \fi
}%
\providecommand \natexlab [1]{#1}%
\providecommand \enquote  [1]{``#1''}%
\providecommand \bibnamefont  [1]{#1}%
\providecommand \bibfnamefont [1]{#1}%
\providecommand \citenamefont [1]{#1}%
\providecommand \href@noop [0]{\@secondoftwo}%
\providecommand \href [0]{\begingroup \@sanitize@url \@href}%
\providecommand \@href[1]{\@@startlink{#1}\@@href}%
\providecommand \@@href[1]{\endgroup#1\@@endlink}%
\providecommand \@sanitize@url [0]{\catcode `\\12\catcode `\$12\catcode `\&12\catcode `\#12\catcode `\^12\catcode `\_12\catcode `\%12\relax}%
\providecommand \@@startlink[1]{}%
\providecommand \@@endlink[0]{}%
\providecommand \url  [0]{\begingroup\@sanitize@url \@url }%
\providecommand \@url [1]{\endgroup\@href {#1}{\urlprefix }}%
\providecommand \urlprefix  [0]{URL }%
\providecommand \Eprint [0]{\href }%
\providecommand \doibase [0]{http://dx.doi.org/}%
\providecommand \selectlanguage [0]{\@gobble}%
\providecommand \bibinfo  [0]{\@secondoftwo}%
\providecommand \bibfield  [0]{\@secondoftwo}%
\providecommand \translation [1]{[#1]}%
\providecommand \BibitemOpen [0]{}%
\providecommand \bibitemStop [0]{}%
\providecommand \bibitemNoStop [0]{.\EOS\space}%
\providecommand \EOS [0]{\spacefactor3000\relax}%
\providecommand \BibitemShut  [1]{\csname bibitem#1\endcsname}%
\let\auto@bib@innerbib\@empty
\bibitem [{\citenamefont {Kirilyuk}, \citenamefont {Kimel},\ and\ \citenamefont {Rasing}(2010)}]{Kirilyuk:2010}%
  \BibitemOpen
  \bibfield  {author} {\bibinfo {author} {\bibfnamefont {A.}~\bibnamefont {Kirilyuk}}, \bibinfo {author} {\bibfnamefont {A.~V.}\ \bibnamefont {Kimel}}, \ and\ \bibinfo {author} {\bibfnamefont {T.}~\bibnamefont {Rasing}},\ }\bibfield  {title} {\enquote {\bibinfo {title} {Ultrafast optical manipulation of magnetic order},}\ }\href {\doibase 10.1103/RevModPhys.82.2731} {\bibfield  {journal} {\bibinfo  {journal} {Rev. Mod. Phys.}\ }\textbf {\bibinfo {volume} {82}},\ \bibinfo {pages} {2731--2784} (\bibinfo {year} {2010})}\BibitemShut {NoStop}%
\bibitem [{\citenamefont {Walowski}\ and\ \citenamefont {Münzenberg}(2016)}]{Walowski:2016}%
  \BibitemOpen
  \bibfield  {author} {\bibinfo {author} {\bibfnamefont {J.}~\bibnamefont {Walowski}}\ and\ \bibinfo {author} {\bibfnamefont {M.}~\bibnamefont {Münzenberg}},\ }\bibfield  {title} {\enquote {\bibinfo {title} {{Perspective: Ultrafast magnetism and THz spintronics}},}\ }\href {\doibase 10.1063/1.4958846} {\bibfield  {journal} {\bibinfo  {journal} {Journal of Applied Physics}\ }\textbf {\bibinfo {volume} {120}},\ \bibinfo {pages} {140901} (\bibinfo {year} {2016})}\BibitemShut {NoStop}%
\bibitem [{\citenamefont {Choi}, \citenamefont {Schleife},\ and\ \citenamefont {Cahill}(2017)}]{Choi:2017}%
  \BibitemOpen
  \bibfield  {author} {\bibinfo {author} {\bibfnamefont {G.-M.}\ \bibnamefont {Choi}}, \bibinfo {author} {\bibfnamefont {A.}~\bibnamefont {Schleife}}, \ and\ \bibinfo {author} {\bibfnamefont {D.~G.}\ \bibnamefont {Cahill}},\ }\bibfield  {title} {\enquote {\bibinfo {title} {Optical-helicity-driven magnetization dynamics in metallic ferromagnets},}\ }\href {\doibase 10.1038/ncomms15085} {\bibfield  {journal} {\bibinfo  {journal} {Nat. Commun.}\ }\textbf {\bibinfo {volume} {8}},\ \bibinfo {pages} {15085} (\bibinfo {year} {2017})}\BibitemShut {NoStop}%
\bibitem [{\citenamefont {Choi}\ \emph {et~al.}(2020{\natexlab{a}})\citenamefont {Choi}, \citenamefont {Oh}, \citenamefont {Lee}, \citenamefont {Lee}, \citenamefont {Kim}, \citenamefont {Lim}, \citenamefont {Min}, \citenamefont {Lee},\ and\ \citenamefont {Lee}}]{Choi:2020}%
  \BibitemOpen
  \bibfield  {author} {\bibinfo {author} {\bibfnamefont {G.-M.}\ \bibnamefont {Choi}}, \bibinfo {author} {\bibfnamefont {J.~H.}\ \bibnamefont {Oh}}, \bibinfo {author} {\bibfnamefont {D.-K.}\ \bibnamefont {Lee}}, \bibinfo {author} {\bibfnamefont {S.-W.}\ \bibnamefont {Lee}}, \bibinfo {author} {\bibfnamefont {K.~W.}\ \bibnamefont {Kim}}, \bibinfo {author} {\bibfnamefont {M.}~\bibnamefont {Lim}}, \bibinfo {author} {\bibfnamefont {B.-C.}\ \bibnamefont {Min}}, \bibinfo {author} {\bibfnamefont {K.-J.}\ \bibnamefont {Lee}}, \ and\ \bibinfo {author} {\bibfnamefont {H.-W.}\ \bibnamefont {Lee}},\ }\bibfield  {title} {\enquote {\bibinfo {title} {Optical spin-orbit torque in heavy metal-ferromagnet heterostructures},}\ }\href {\doibase 10.1038/s41467-020-15247-3} {\bibfield  {journal} {\bibinfo  {journal} {Nat. Commun.}\ }\textbf {\bibinfo {volume} {11}},\ \bibinfo {pages} {1482} (\bibinfo {year} {2020}{\natexlab{a}})}\BibitemShut {NoStop}%
\bibitem [{\citenamefont {Iihama}, \citenamefont {Ishibashi},\ and\ \citenamefont {Mizukami}(2022)}]{Iihama:2022}%
  \BibitemOpen
  \bibfield  {author} {\bibinfo {author} {\bibfnamefont {S.}~\bibnamefont {Iihama}}, \bibinfo {author} {\bibfnamefont {K.}~\bibnamefont {Ishibashi}}, \ and\ \bibinfo {author} {\bibfnamefont {S.}~\bibnamefont {Mizukami}},\ }\bibfield  {title} {\enquote {\bibinfo {title} {{Photon spin angular momentum driven magnetization dynamics in ferromagnet/heavy metal bilayers}},}\ }\href {\doibase 10.1063/5.0073409} {\bibfield  {journal} {\bibinfo  {journal} {J. Appl. Phys.}\ }\textbf {\bibinfo {volume} {131}},\ \bibinfo {pages} {023901} (\bibinfo {year} {2022})}\BibitemShut {NoStop}%
\bibitem [{\citenamefont {Malinowski}\ \emph {et~al.}(2008)\citenamefont {Malinowski}, \citenamefont {Dalla~Longa}, \citenamefont {Rietjens}, \citenamefont {Paluskar}, \citenamefont {Huijink}, \citenamefont {Swagten},\ and\ \citenamefont {Koopmans}}]{Malinowski:2008}%
  \BibitemOpen
  \bibfield  {author} {\bibinfo {author} {\bibfnamefont {G.}~\bibnamefont {Malinowski}}, \bibinfo {author} {\bibfnamefont {F.}~\bibnamefont {Dalla~Longa}}, \bibinfo {author} {\bibfnamefont {J.~H.~H.}\ \bibnamefont {Rietjens}}, \bibinfo {author} {\bibfnamefont {P.~V.}\ \bibnamefont {Paluskar}}, \bibinfo {author} {\bibfnamefont {R.}~\bibnamefont {Huijink}}, \bibinfo {author} {\bibfnamefont {H.~J.~M.}\ \bibnamefont {Swagten}}, \ and\ \bibinfo {author} {\bibfnamefont {B.}~\bibnamefont {Koopmans}},\ }\bibfield  {title} {\enquote {\bibinfo {title} {Control of speed and efficiency of ultrafast demagnetization by direct transfer of spin angular momentum},}\ }\href {\doibase 10.1038/nphys1092} {\bibfield  {journal} {\bibinfo  {journal} {Nat. Phys.}\ }\textbf {\bibinfo {volume} {4}},\ \bibinfo {pages} {855} (\bibinfo {year} {2008})}\BibitemShut {NoStop}%
\bibitem [{\citenamefont {Schellekens}\ \emph {et~al.}(2014)\citenamefont {Schellekens}, \citenamefont {Kuiper}, \citenamefont {de~Wit},\ and\ \citenamefont {Koopmans}}]{Schellekens:2014b}%
  \BibitemOpen
  \bibfield  {author} {\bibinfo {author} {\bibfnamefont {A.~J.}\ \bibnamefont {Schellekens}}, \bibinfo {author} {\bibfnamefont {K.~C.}\ \bibnamefont {Kuiper}}, \bibinfo {author} {\bibfnamefont {R.~R. J.~C.}\ \bibnamefont {de~Wit}}, \ and\ \bibinfo {author} {\bibfnamefont {B.}~\bibnamefont {Koopmans}},\ }\bibfield  {title} {\enquote {\bibinfo {title} {Ultrafast spin-transfer torque driven by femtosecond pulsed-laser excitation},}\ }\href {\doibase 10.1038/ncomms5333} {\bibfield  {journal} {\bibinfo  {journal} {Nat. Comm.}\ }\textbf {\bibinfo {volume} {5}},\ \bibinfo {pages} {4333} (\bibinfo {year} {2014})}\BibitemShut {NoStop}%
\bibitem [{\citenamefont {van Hees}\ \emph {et~al.}(2020)\citenamefont {van Hees}, \citenamefont {van~de Meugheuvel}, \citenamefont {Koopmans},\ and\ \citenamefont {Lavrijsen}}]{vanHees:2020}%
  \BibitemOpen
  \bibfield  {author} {\bibinfo {author} {\bibfnamefont {Y.~L.~W.}\ \bibnamefont {van Hees}}, \bibinfo {author} {\bibfnamefont {P.}~\bibnamefont {van~de Meugheuvel}}, \bibinfo {author} {\bibfnamefont {B.}~\bibnamefont {Koopmans}}, \ and\ \bibinfo {author} {\bibfnamefont {R.}~\bibnamefont {Lavrijsen}},\ }\bibfield  {title} {\enquote {\bibinfo {title} {Deterministic all-optical magnetization writing facilitated by non-local transfer of spin angular momentum},}\ }\href {\doibase 10.1038/s41467-020-17676-6} {\bibfield  {journal} {\bibinfo  {journal} {Nat. Comm.}\ }\textbf {\bibinfo {volume} {11}},\ \bibinfo {pages} {3835} (\bibinfo {year} {2020})}\BibitemShut {NoStop}%
\bibitem [{\citenamefont {Beens}, \citenamefont {Duine},\ and\ \citenamefont {Koopmans}(2020)}]{Beens:2020}%
  \BibitemOpen
  \bibfield  {author} {\bibinfo {author} {\bibfnamefont {M.}~\bibnamefont {Beens}}, \bibinfo {author} {\bibfnamefont {R.~A.}\ \bibnamefont {Duine}}, \ and\ \bibinfo {author} {\bibfnamefont {B.}~\bibnamefont {Koopmans}},\ }\bibfield  {title} {\enquote {\bibinfo {title} {$s\text{\ensuremath{-}}d$ model for local and nonlocal spin dynamics in laser-excited magnetic heterostructures},}\ }\href {\doibase 10.1103/PhysRevB.102.054442} {\bibfield  {journal} {\bibinfo  {journal} {Phys.~Rev.~B}\ }\textbf {\bibinfo {volume} {102}},\ \bibinfo {pages} {054442} (\bibinfo {year} {2020})}\BibitemShut {NoStop}%
\bibitem [{\citenamefont {Choi}\ \emph {et~al.}(2020{\natexlab{b}})\citenamefont {Choi}, \citenamefont {Lee}, \citenamefont {Lee},\ and\ \citenamefont {Lee}}]{Choi:2020b}%
  \BibitemOpen
  \bibfield  {author} {\bibinfo {author} {\bibfnamefont {G.-M.}\ \bibnamefont {Choi}}, \bibinfo {author} {\bibfnamefont {D.-K.}\ \bibnamefont {Lee}}, \bibinfo {author} {\bibfnamefont {K.-J.}\ \bibnamefont {Lee}}, \ and\ \bibinfo {author} {\bibfnamefont {H.-W.}\ \bibnamefont {Lee}},\ }\bibfield  {title} {\enquote {\bibinfo {title} {Coherent spin waves driven by optical spin-orbit torque},}\ }\href {\doibase 10.1103/PhysRevB.102.014437} {\bibfield  {journal} {\bibinfo  {journal} {Phys. Rev. B}\ }\textbf {\bibinfo {volume} {102}},\ \bibinfo {pages} {014437} (\bibinfo {year} {2020}{\natexlab{b}})}\BibitemShut {NoStop}%
\bibitem [{\citenamefont {Keffer}\ and\ \citenamefont {Kittel}(1952)}]{Keffer:1952}%
  \BibitemOpen
  \bibfield  {author} {\bibinfo {author} {\bibfnamefont {F.}~\bibnamefont {Keffer}}\ and\ \bibinfo {author} {\bibfnamefont {C.}~\bibnamefont {Kittel}},\ }\bibfield  {title} {\enquote {\bibinfo {title} {Theory of antiferromagnetic resonance},}\ }\href {\doibase 10.1103/PhysRev.85.329} {\bibfield  {journal} {\bibinfo  {journal} {Phys. Rev.}\ }\textbf {\bibinfo {volume} {85}},\ \bibinfo {pages} {329} (\bibinfo {year} {1952})}\BibitemShut {NoStop}%
\bibitem [{\citenamefont {Hintermayr}, \citenamefont {van Hees},\ and\ \citenamefont {Koopmans}(2023)}]{Hintermayr:2023}%
  \BibitemOpen
  \bibfield  {author} {\bibinfo {author} {\bibfnamefont {J.}~\bibnamefont {Hintermayr}}, \bibinfo {author} {\bibfnamefont {Y.~L.~W.}\ \bibnamefont {van Hees}}, \ and\ \bibinfo {author} {\bibfnamefont {B.}~\bibnamefont {Koopmans}},\ }\bibfield  {title} {\enquote {\bibinfo {title} {Exploring terahertz-scale exchange resonances in synthetic ferrimagnets with ultrashort optically induced spin currents},}\ }\href {\doibase 10.1103/PhysRevB.108.024401} {\bibfield  {journal} {\bibinfo  {journal} {Phys. Rev. B}\ }\textbf {\bibinfo {volume} {108}},\ \bibinfo {pages} {024401} (\bibinfo {year} {2023})}\BibitemShut {NoStop}%
\bibitem [{\citenamefont {Blank}\ \emph {et~al.}(2021)\citenamefont {Blank}, \citenamefont {Grishunin}, \citenamefont {Mashkovich}, \citenamefont {Logunov}, \citenamefont {Zvezdin},\ and\ \citenamefont {Kimel}}]{Blank:2021}%
  \BibitemOpen
  \bibfield  {author} {\bibinfo {author} {\bibfnamefont {T.~G.~H.}\ \bibnamefont {Blank}}, \bibinfo {author} {\bibfnamefont {K.~A.}\ \bibnamefont {Grishunin}}, \bibinfo {author} {\bibfnamefont {E.~A.}\ \bibnamefont {Mashkovich}}, \bibinfo {author} {\bibfnamefont {M.~V.}\ \bibnamefont {Logunov}}, \bibinfo {author} {\bibfnamefont {A.~K.}\ \bibnamefont {Zvezdin}}, \ and\ \bibinfo {author} {\bibfnamefont {A.~V.}\ \bibnamefont {Kimel}},\ }\bibfield  {title} {\enquote {\bibinfo {title} {{THz}-scale field-induced spin dynamics in ferrimagnetic iron garnets},}\ }\href {\doibase 10.1103/PhysRevLett.127.037203} {\bibfield  {journal} {\bibinfo  {journal} {Phys. Rev. Lett.}\ }\textbf {\bibinfo {volume} {127}},\ \bibinfo {pages} {037203} (\bibinfo {year} {2021})}\BibitemShut {NoStop}%
\bibitem [{\citenamefont {Stanciu}\ \emph {et~al.}(2006)\citenamefont {Stanciu}, \citenamefont {Kimel}, \citenamefont {Hansteen}, \citenamefont {Tsukamoto}, \citenamefont {Itoh}, \citenamefont {Kiriliyuk},\ and\ \citenamefont {Rasing}}]{Stanciu:2006}%
  \BibitemOpen
  \bibfield  {author} {\bibinfo {author} {\bibfnamefont {C.~D.}\ \bibnamefont {Stanciu}}, \bibinfo {author} {\bibfnamefont {A.~V.}\ \bibnamefont {Kimel}}, \bibinfo {author} {\bibfnamefont {F.}~\bibnamefont {Hansteen}}, \bibinfo {author} {\bibfnamefont {A.}~\bibnamefont {Tsukamoto}}, \bibinfo {author} {\bibfnamefont {A.}~\bibnamefont {Itoh}}, \bibinfo {author} {\bibfnamefont {A.}~\bibnamefont {Kiriliyuk}}, \ and\ \bibinfo {author} {\bibfnamefont {T.}~\bibnamefont {Rasing}},\ }\bibfield  {title} {\enquote {\bibinfo {title} {Ultrafast spin dynamics across compensation points in ferrimagnetic $\mathrm{GdFeCo}$: The role of angular momentum compensation},}\ }\href {\doibase 10.1103/PhysRevB.73.220402} {\bibfield  {journal} {\bibinfo  {journal} {Phys. Rev. B}\ }\textbf {\bibinfo {volume} {73}},\ \bibinfo {pages} {220402(R)} (\bibinfo {year} {2006})}\BibitemShut {NoStop}%
\bibitem [{\citenamefont {Mekonnen}\ \emph {et~al.}(2011)\citenamefont {Mekonnen}, \citenamefont {Cormier}, \citenamefont {Kimel}, \citenamefont {Kirilyuk}, \citenamefont {Hrabec}, \citenamefont {Ranno},\ and\ \citenamefont {Rasing}}]{Mekonnen:2011}%
  \BibitemOpen
  \bibfield  {author} {\bibinfo {author} {\bibfnamefont {A.}~\bibnamefont {Mekonnen}}, \bibinfo {author} {\bibfnamefont {M.}~\bibnamefont {Cormier}}, \bibinfo {author} {\bibfnamefont {A.~V.}\ \bibnamefont {Kimel}}, \bibinfo {author} {\bibfnamefont {A.}~\bibnamefont {Kirilyuk}}, \bibinfo {author} {\bibfnamefont {A.}~\bibnamefont {Hrabec}}, \bibinfo {author} {\bibfnamefont {L.}~\bibnamefont {Ranno}}, \ and\ \bibinfo {author} {\bibfnamefont {T.}~\bibnamefont {Rasing}},\ }\bibfield  {title} {\enquote {\bibinfo {title} {Femtosecond laser excitation of spin resonances in amorphous ferrimagnetic {Gd\textsubscript{1-\textit{x}}Co\textsubscript{\textit{x}}} alloys},}\ }\href {\doibase 10.1103/PhysRevLett.107.117202} {\bibfield  {journal} {\bibinfo  {journal} {Phys. Rev. Lett.}\ }\textbf {\bibinfo {volume} {107}},\ \bibinfo {pages} {117202} (\bibinfo {year} {2011})}\BibitemShut {NoStop}%
\bibitem [{\citenamefont {Haltz}\ \emph {et~al.}(2022)\citenamefont {Haltz}, \citenamefont {Sampaio}, \citenamefont {Krishnia}, \citenamefont {Berges}, \citenamefont {Weil}, \citenamefont {Mougin},\ and\ \citenamefont {Thiaville}}]{Haltz:2022}%
  \BibitemOpen
  \bibfield  {author} {\bibinfo {author} {\bibfnamefont {E.}~\bibnamefont {Haltz}}, \bibinfo {author} {\bibfnamefont {J.~a.}\ \bibnamefont {Sampaio}}, \bibinfo {author} {\bibfnamefont {S.}~\bibnamefont {Krishnia}}, \bibinfo {author} {\bibfnamefont {L.}~\bibnamefont {Berges}}, \bibinfo {author} {\bibfnamefont {R.}~\bibnamefont {Weil}}, \bibinfo {author} {\bibfnamefont {A.}~\bibnamefont {Mougin}}, \ and\ \bibinfo {author} {\bibfnamefont {A.}~\bibnamefont {Thiaville}},\ }\bibfield  {title} {\enquote {\bibinfo {title} {Quantitative analysis of spin wave dynamics in ferrimagnets across compensation points},}\ }\href {\doibase 10.1103/PhysRevB.105.104414} {\bibfield  {journal} {\bibinfo  {journal} {Phys. Rev. B}\ }\textbf {\bibinfo {volume} {105}},\ \bibinfo {pages} {104414} (\bibinfo {year} {2022})}\BibitemShut {NoStop}%
\bibitem [{\citenamefont {Erskine}\ and\ \citenamefont {Stern}(1973)}]{Erskine:1973}%
  \BibitemOpen
  \bibfield  {author} {\bibinfo {author} {\bibfnamefont {J.~L.}\ \bibnamefont {Erskine}}\ and\ \bibinfo {author} {\bibfnamefont {E.~A.}\ \bibnamefont {Stern}},\ }\bibfield  {title} {\enquote {\bibinfo {title} {Magneto-optic kerr effects in gadolinium},}\ }\href {\doibase 10.1103/PhysRevB.8.1239} {\bibfield  {journal} {\bibinfo  {journal} {Phys. Rev. B}\ }\textbf {\bibinfo {volume} {8}},\ \bibinfo {pages} {1239--1255} (\bibinfo {year} {1973})}\BibitemShut {NoStop}%
\bibitem [{\citenamefont {Eichler}, \citenamefont {Langhans},\ and\ \citenamefont {Massmann}(1984)}]{Eichler:1984}%
  \BibitemOpen
  \bibfield  {author} {\bibinfo {author} {\bibfnamefont {H.}~\bibnamefont {Eichler}}, \bibinfo {author} {\bibfnamefont {D.}~\bibnamefont {Langhans}}, \ and\ \bibinfo {author} {\bibfnamefont {F.}~\bibnamefont {Massmann}},\ }\bibfield  {title} {\enquote {\bibinfo {title} {Coherence peaks in picosecond sampling experiments},}\ }\href {\doibase 10.1016/0030-4018(84)90147-0} {\bibfield  {journal} {\bibinfo  {journal} {Opt. Commun.}\ }\textbf {\bibinfo {volume} {50}},\ \bibinfo {pages} {117} (\bibinfo {year} {1984})}\BibitemShut {NoStop}%
\bibitem [{\citenamefont {Luo}\ \emph {et~al.}(2009)\citenamefont {Luo}, \citenamefont {Wang}, \citenamefont {Chen}, \citenamefont {Shih},\ and\ \citenamefont {Kobayashi}}]{Luo:2009}%
  \BibitemOpen
  \bibfield  {author} {\bibinfo {author} {\bibfnamefont {C.~W.}\ \bibnamefont {Luo}}, \bibinfo {author} {\bibfnamefont {Y.~T.}\ \bibnamefont {Wang}}, \bibinfo {author} {\bibfnamefont {F.~W.}\ \bibnamefont {Chen}}, \bibinfo {author} {\bibfnamefont {H.~C.}\ \bibnamefont {Shih}}, \ and\ \bibinfo {author} {\bibfnamefont {T.}~\bibnamefont {Kobayashi}},\ }\bibfield  {title} {\enquote {\bibinfo {title} {Eliminate coherence spike in reflection-type pump-probe measurements},}\ }\href {\doibase 10.1364/OE.17.011321} {\bibfield  {journal} {\bibinfo  {journal} {Opt. Express}\ }\textbf {\bibinfo {volume} {17}},\ \bibinfo {pages} {11321} (\bibinfo {year} {2009})}\BibitemShut {NoStop}%
\bibitem [{\citenamefont {Freimuth}, \citenamefont {Bl\"ugel},\ and\ \citenamefont {Mokrousov}(2016)}]{Freimuth:2016}%
  \BibitemOpen
  \bibfield  {author} {\bibinfo {author} {\bibfnamefont {F.}~\bibnamefont {Freimuth}}, \bibinfo {author} {\bibfnamefont {S.}~\bibnamefont {Bl\"ugel}}, \ and\ \bibinfo {author} {\bibfnamefont {Y.}~\bibnamefont {Mokrousov}},\ }\bibfield  {title} {\enquote {\bibinfo {title} {Laser-induced torques in metallic ferromagnets},}\ }\href {\doibase 10.1103/PhysRevB.94.144432} {\bibfield  {journal} {\bibinfo  {journal} {Phys. Rev. B}\ }\textbf {\bibinfo {volume} {94}},\ \bibinfo {pages} {144432} (\bibinfo {year} {2016})}\BibitemShut {NoStop}%
\bibitem [{\citenamefont {Schlickeiser}\ \emph {et~al.}(2012)\citenamefont {Schlickeiser}, \citenamefont {Atxitia}, \citenamefont {Wienholdt}, \citenamefont {Hinzke}, \citenamefont {Chubykalo-Fesenko},\ and\ \citenamefont {Nowak}}]{Schlickeiser:2012}%
  \BibitemOpen
  \bibfield  {author} {\bibinfo {author} {\bibfnamefont {F.}~\bibnamefont {Schlickeiser}}, \bibinfo {author} {\bibfnamefont {U.}~\bibnamefont {Atxitia}}, \bibinfo {author} {\bibfnamefont {S.}~\bibnamefont {Wienholdt}}, \bibinfo {author} {\bibfnamefont {D.}~\bibnamefont {Hinzke}}, \bibinfo {author} {\bibfnamefont {O.}~\bibnamefont {Chubykalo-Fesenko}}, \ and\ \bibinfo {author} {\bibfnamefont {U.}~\bibnamefont {Nowak}},\ }\bibfield  {title} {\enquote {\bibinfo {title} {Temperature dependence of the frequencies and effective damping parameters of ferrimagnetic resonance},}\ }\href {\doibase 10.1103/PhysRevB.86.214416} {\bibfield  {journal} {\bibinfo  {journal} {Phys. Rev. B}\ }\textbf {\bibinfo {volume} {86}},\ \bibinfo {pages} {214416} (\bibinfo {year} {2012})}\BibitemShut {NoStop}%
\bibitem [{\citenamefont {Lalieu}, \citenamefont {Helgers},\ and\ \citenamefont {Koopmans}(2017)}]{Lalieu:2017b}%
  \BibitemOpen
  \bibfield  {author} {\bibinfo {author} {\bibfnamefont {M.~L.~M.}\ \bibnamefont {Lalieu}}, \bibinfo {author} {\bibfnamefont {P.~L.~J.}\ \bibnamefont {Helgers}}, \ and\ \bibinfo {author} {\bibfnamefont {B.}~\bibnamefont {Koopmans}},\ }\bibfield  {title} {\enquote {\bibinfo {title} {Absorption and generation of femtosecond laser-pulse excited spin currents in noncollinear magnetic bilayers},}\ }\href {\doibase 10.1103/PhysRevB.96.014417} {\bibfield  {journal} {\bibinfo  {journal} {Phys. Rev. B}\ }\textbf {\bibinfo {volume} {96}},\ \bibinfo {pages} {014417} (\bibinfo {year} {2017})}\BibitemShut {NoStop}%
\bibitem [{\citenamefont {Battiato}, \citenamefont {Carva},\ and\ \citenamefont {Oppeneer}(2010)}]{Battiato:2010}%
  \BibitemOpen
  \bibfield  {author} {\bibinfo {author} {\bibfnamefont {M.}~\bibnamefont {Battiato}}, \bibinfo {author} {\bibfnamefont {K.}~\bibnamefont {Carva}}, \ and\ \bibinfo {author} {\bibfnamefont {P.~M.}\ \bibnamefont {Oppeneer}},\ }\bibfield  {title} {\enquote {\bibinfo {title} {Superdiffusive spin transport as a mechanism of ultrafast demagnetization},}\ }\href {\doibase 10.1103/PhysRevLett.105.027203} {\bibfield  {journal} {\bibinfo  {journal} {Phys.~Rev.~Lett.}\ }\textbf {\bibinfo {volume} {105}},\ \bibinfo {pages} {027203} (\bibinfo {year} {2010})}\BibitemShut {NoStop}%
\bibitem [{\citenamefont {Battiato}, \citenamefont {Carva},\ and\ \citenamefont {Oppeneer}(2012)}]{Battiato:2012}%
  \BibitemOpen
  \bibfield  {author} {\bibinfo {author} {\bibfnamefont {M.}~\bibnamefont {Battiato}}, \bibinfo {author} {\bibfnamefont {K.}~\bibnamefont {Carva}}, \ and\ \bibinfo {author} {\bibfnamefont {P.~M.}\ \bibnamefont {Oppeneer}},\ }\bibfield  {title} {\enquote {\bibinfo {title} {Theory of laser-induced ultrafast superdiffusive spin transport in layered heterostructures},}\ }\href {\doibase 10.1103/PhysRevB.86.024404} {\bibfield  {journal} {\bibinfo  {journal} {Phys. Rev. B}\ }\textbf {\bibinfo {volume} {86}},\ \bibinfo {pages} {024404} (\bibinfo {year} {2012})}\BibitemShut {NoStop}%
\bibitem [{\citenamefont {Lichtenberg}\ \emph {et~al.}(2022)\citenamefont {Lichtenberg}, \citenamefont {Beens}, \citenamefont {Jansen}, \citenamefont {Koopmans},\ and\ \citenamefont {Duine}}]{Lichtenberg:2022}%
  \BibitemOpen
  \bibfield  {author} {\bibinfo {author} {\bibfnamefont {T.}~\bibnamefont {Lichtenberg}}, \bibinfo {author} {\bibfnamefont {M.}~\bibnamefont {Beens}}, \bibinfo {author} {\bibfnamefont {M.~H.}\ \bibnamefont {Jansen}}, \bibinfo {author} {\bibfnamefont {B.}~\bibnamefont {Koopmans}}, \ and\ \bibinfo {author} {\bibfnamefont {R.~A.}\ \bibnamefont {Duine}},\ }\bibfield  {title} {\enquote {\bibinfo {title} {Probing optically induced spin currents using terahertz spin waves in noncollinear magnetic bilayers},}\ }\href {\doibase 10.1103/PhysRevB.105.144416} {\bibfield  {journal} {\bibinfo  {journal} {Phys. Rev. B}\ }\textbf {\bibinfo {volume} {105}},\ \bibinfo {pages} {144416} (\bibinfo {year} {2022})}\BibitemShut {NoStop}%
\bibitem [{\citenamefont {Gupta}\ \emph {et~al.}(2023)\citenamefont {Gupta}, \citenamefont {Cosco}, \citenamefont {Malik}, \citenamefont {Chen}, \citenamefont {Saha}, \citenamefont {Ghosh}, \citenamefont {Pohlmann}, \citenamefont {Mardegan}, \citenamefont {Francoual}, \citenamefont {Stefanuik}, \citenamefont {S\"oderstr\"om}, \citenamefont {Sanyal}, \citenamefont {Karis}, \citenamefont {Svedlindh}, \citenamefont {Oppeneer},\ and\ \citenamefont {Knut}}]{Gupta:2023}%
  \BibitemOpen
  \bibfield  {author} {\bibinfo {author} {\bibfnamefont {R.}~\bibnamefont {Gupta}}, \bibinfo {author} {\bibfnamefont {F.}~\bibnamefont {Cosco}}, \bibinfo {author} {\bibfnamefont {R.~S.}\ \bibnamefont {Malik}}, \bibinfo {author} {\bibfnamefont {X.}~\bibnamefont {Chen}}, \bibinfo {author} {\bibfnamefont {S.}~\bibnamefont {Saha}}, \bibinfo {author} {\bibfnamefont {A.}~\bibnamefont {Ghosh}}, \bibinfo {author} {\bibfnamefont {T.}~\bibnamefont {Pohlmann}}, \bibinfo {author} {\bibfnamefont {J.~R.~L.}\ \bibnamefont {Mardegan}}, \bibinfo {author} {\bibfnamefont {S.}~\bibnamefont {Francoual}}, \bibinfo {author} {\bibfnamefont {R.}~\bibnamefont {Stefanuik}}, \bibinfo {author} {\bibfnamefont {J.}~\bibnamefont {S\"oderstr\"om}}, \bibinfo {author} {\bibfnamefont {B.}~\bibnamefont {Sanyal}}, \bibinfo {author} {\bibfnamefont {O.}~\bibnamefont {Karis}}, \bibinfo {author} {\bibfnamefont {P.}~\bibnamefont {Svedlindh}}, \bibinfo {author} {\bibfnamefont {P.~M.}\ \bibnamefont {Oppeneer}}, \ and\ \bibinfo {author} {\bibfnamefont
  {R.}~\bibnamefont {Knut}},\ }\bibfield  {title} {\enquote {\bibinfo {title} {Element-resolved evidence of superdiffusive spin current arising from ultrafast demagnetization process},}\ }\href {\doibase 10.1103/PhysRevB.108.064427} {\bibfield  {journal} {\bibinfo  {journal} {Phys. Rev. B}\ }\textbf {\bibinfo {volume} {108}},\ \bibinfo {pages} {064427} (\bibinfo {year} {2023})}\BibitemShut {NoStop}%
\bibitem [{\citenamefont {Hamamera}\ \emph {et~al.}(2024)\citenamefont {Hamamera}, \citenamefont {Guimarães}, \citenamefont {Dos Santos~Dias},\ and\ \citenamefont {Lounis}}]{Hamamera:2024}%
  \BibitemOpen
  \bibfield  {author} {\bibinfo {author} {\bibfnamefont {H.}~\bibnamefont {Hamamera}}, \bibinfo {author} {\bibfnamefont {F.~S.}\ \bibnamefont {Guimarães}}, \bibinfo {author} {\bibfnamefont {M.}~\bibnamefont {Dos Santos~Dias}}, \ and\ \bibinfo {author} {\bibfnamefont {S.}~\bibnamefont {Lounis}},\ }\bibfield  {title} {\enquote {\bibinfo {title} {Ultrafast light-induced magnetization in non-magnetic films: from orbital and spin hall phenomena to the inverse faraday effect},}\ }\href {\doibase 10.3389/fphy.2024.1354870} {\bibfield  {journal} {\bibinfo  {journal} {Front.~Phys.}\ }\textbf {\bibinfo {volume} {12}} (\bibinfo {year} {2024}),\ 10.3389/fphy.2024.1354870}\BibitemShut {NoStop}%
\bibitem [{\citenamefont {Busch}\ \emph {et~al.}(2023)\citenamefont {Busch}, \citenamefont {Ziolkowski}, \citenamefont {Mertig},\ and\ \citenamefont {Henk}}]{Busch:2023}%
  \BibitemOpen
  \bibfield  {author} {\bibinfo {author} {\bibfnamefont {O.}~\bibnamefont {Busch}}, \bibinfo {author} {\bibfnamefont {F.}~\bibnamefont {Ziolkowski}}, \bibinfo {author} {\bibfnamefont {I.}~\bibnamefont {Mertig}}, \ and\ \bibinfo {author} {\bibfnamefont {J.}~\bibnamefont {Henk}},\ }\bibfield  {title} {\enquote {\bibinfo {title} {Ultrafast dynamics of orbital angular momentum of electrons induced by femtosecond laser pulses: Generation and transfer across interfaces},}\ }\href {\doibase 10.1103/PhysRevB.108.104408} {\bibfield  {journal} {\bibinfo  {journal} {Phys. Rev. B}\ }\textbf {\bibinfo {volume} {108}},\ \bibinfo {pages} {104408} (\bibinfo {year} {2023})}\BibitemShut {NoStop}%
\bibitem [{\citenamefont {Sala}\ and\ \citenamefont {Gambardella}(2022)}]{Sala:2022}%
  \BibitemOpen
  \bibfield  {author} {\bibinfo {author} {\bibfnamefont {G.}~\bibnamefont {Sala}}\ and\ \bibinfo {author} {\bibfnamefont {P.}~\bibnamefont {Gambardella}},\ }\bibfield  {title} {\enquote {\bibinfo {title} {Giant orbital hall effect and orbital-to-spin conversion in $3d$, $5d$, and $4f$ metallic heterostructures},}\ }\href {\doibase 10.1103/PhysRevResearch.4.033037} {\bibfield  {journal} {\bibinfo  {journal} {Phys. Rev. Res.}\ }\textbf {\bibinfo {volume} {4}},\ \bibinfo {pages} {033037} (\bibinfo {year} {2022})}\BibitemShut {NoStop}%
\end{thebibliography}%

\end{document}